\documentclass[letterpaper, 10 pt, conference]{ieeeconf}  
\usepackage{amssymb}

\usepackage{algorithm}
\usepackage[noend]{algpseudocode}
\usepackage{algorithm}
\usepackage{pifont}
\usepackage{amsfonts}
\usepackage{kantlipsum}       
\usepackage{mwe}  
\usepackage{float}                 
\usepackage{graphicx}
\usepackage{caption}
\usepackage{hyperref}
\usepackage{pifont}
\usepackage{subcaption}
\usepackage{textcomp}
\usepackage[english]{babel}
\usepackage{multicol}
\usepackage[noadjust]{cite}
\usepackage{textcomp}
\usepackage{mathtools}
\newtheorem{theorem}{Theorem}
\newtheorem{lemma}{Lemma}
\newtheorem{proposition}{Proposition}

\newtheorem{remark}{Remark}
\usepackage{booktabs}  
\usepackage{amsmath}   

\newcommand{\norm}[1]{\|#1\|}
\newcommand{\Touter}{T_{\text{outer}}}
\newcommand{\Tinner}{T_{\text{inner}}}
\newcommand{\oscK}{\operatorname{osc}(\log K)}
\newcommand{\dH}{d_H}  
\newcommand{\qed}{\hfill$\square$}
\newcommand{\proj}[1]{#1/\langle\mathbf 1,#1\rangle}

\usepackage[noadjust]{cite}                                                        

\IEEEoverridecommandlockouts                              
\title{\LARGE \bf
Geometry-Aware Decentralized Sinkhorn for Wasserstein Barycenters
}

\author{Ali Baheri and Alireza Vahid
\thanks{Ali Baheri is with the Mechanical Engineering Department, 
Alireza Vahid is with the Electrical and Microelectronic Engineering Department, 
both at Rochester Institute of Technology, Rochester, NY 14623. 
Emails: {\tt\small akbeme@rit.edu, arveme@rit.edu}.}}

\DeclareMathOperator{\softmax}{softmax}
\begin{document}

\maketitle
\thispagestyle{empty}
\pagestyle{empty}

\begin{abstract}

Distributed systems require fusing heterogeneous local probability distributions into a global summary over sparse and unreliable communication networks. Traditional consensus algorithms, which average distributions in Euclidean space, ignore their inherent geometric structure, leading to misleading results. Wasserstein barycenters offer a geometry-aware alternative by minimizing optimal transport costs, but their entropic approximations via the Sinkhorn algorithm typically require centralized coordination.
This paper proposes a fully decentralized Sinkhorn algorithm that reformulates the centralized geometric mean as an arithmetic average in the log-domain, enabling approximation through local gossip protocols. Agents exchange log-messages with neighbors, interleaving consensus phases with local updates to mimic centralized iterations without a coordinator. To optimize bandwidth, we integrate event-triggered transmissions and b-bit quantization, providing tunable trade-offs between accuracy and communication while accommodating asynchrony and packet loss. Under mild assumptions, we prove convergence to a neighborhood of the centralized entropic barycenter, with bias linearly dependent on consensus tolerance, trigger threshold, and quantization error. Complexity scales near-linearly with network size. Simulations confirm near-centralized accuracy with significantly fewer messages across various topologies and conditions.

\end{abstract}


\section{INTRODUCTION}

Distributed systems increasingly require fusing heterogeneous local probability distributions into network-wide summaries over sparse, unreliable links~\cite{OlfatiSaber2007Proceedings,mcmahan2017communication}. Applications range from autonomous robots sharing target beliefs to sensor networks aggregating measurements to edge devices collaborating on models without sharing raw data. Classical consensus algorithms treat distributions as Euclidean vectors and compute arithmetic averages~\cite{boyd2006gossip}, fundamentally ignoring geometric structure. Naive averaging can produce misleading results: when two robots believe a target is at opposite locations, their arithmetic mean suggests the midpoint, potentially an impossible location. The core issue is that probability distributions live on a curved manifold, and the cost of transforming one into another depends on the underlying geometry. Optimal transport addresses this by measuring distances via the minimum cost of moving probability mass~\cite{villani2009ot}, and Wasserstein barycenters extend this to geometry-aware averaging that preserves distribution shapes and respects spatial relationships.

Entropic regularization enables practical computation via the Sinkhorn algorithm~\cite{cuturi2013sinkhorn}, which alternates between local scaling operations and a global geometric mean. However, the geometric mean step requires centralized aggregation at every iteration. Existing decentralized and federated approaches typically compute Wasserstein barycenters through consensus-style distributed computation, accelerated primal--dual stochastic optimization, quantized distributed optimization, time-varying-network convex optimization, or server-orchestrated federated protocols~\cite{uribe2018distributed,dvurechensky2018decentralize,krawtschenko2020distributed,yufereva2024decentralized,rakotomamonjy2024federated,wei2025personalized}. These designs either exchange dense dual/model information through general optimization routines or retain a coordinator, and they do not implement the Sinkhorn geometric-mean step as a peer-to-peer log-average. None provides the particular event-triggered, quantized log-gossip mechanism developed here for trading barycenter accuracy against bandwidth.

Our key idea is that the geometric mean can be reformulated as an arithmetic average in log-space, enabling fully distributed computation via neighbor-to-neighbor gossip~\cite{boyd2006gossip}. To address bandwidth constraints, we introduce event-triggered transmission~\cite{tabuada2007event} and adaptive quantization~\cite{aysal2008distributed}, providing explicit controls for trading accuracy against communication cost while supporting asynchrony and packet loss.

\noindent \textbf{Contributions.} \emph{Algorithmically}, we develop a fully decentralized Sinkhorn algorithm that eliminates central coordination by reformulating geometric means as distributed arithmetic consensus in log-space. \emph{Communication-wise}, we design event-triggered and quantized protocols that reduce bandwidth by orders of magnitude while preserving convergence guaranties. \emph{Theoretically}, we establish convergence to the centralized solution with explicit bounds on how consensus tolerance, trigger thresholds, and quantization precision affect accuracy.

\vspace{-3 mm}
\section{Related Work}

\noindent \textbf{Optimal Transport and Wasserstein Barycenters}
Optimal transport is a framework for comparing probability measures with deep geometric structure. Villani's work covers theoretical foundations~\cite{villani2009ot}, while Peyré and Cuturi survey computational aspects~\cite{peyre2019cot}. Wasserstein barycenters, formalized by Agueh and Carlier~\cite{agueh2011barycenters}, extend Fréchet means to the space of probability measures. Optimal transport has found widespread applications across machine learning~\cite{torres2021survey}, computer vision~\cite{solomon2015convolutional}, and reinforcement learning~\cite{baheri2023risk}.

\noindent \textbf{Centralized Barycenter Algorithms}
Entropic regularization~\cite{cuturi2013sinkhorn} enabled practical barycenter computation using the Sinkhorn algorithm and its Bregman projection form~\cite{benamou2015ibp}. Fast variants~\cite{cuturi2014barycenters} provide strong accuracy-complexity trade-offs in centralized settings. Advanced techniques include log-domain stabilization for numerical robustness~\cite{schmitzer2019stabilized}, convolutional kernels for geometric domains~\cite{solomon2015convolutional}, and sliced variants for high dimensions~\cite{bonneel2015sliced}. These methods assume centralized coordination or star-topology parallelism, communicating full histograms or dual potentials at each iteration.


\noindent \textbf{Decentralized Wasserstein Barycenter Computation}
Existing decentralized approaches include consensus-style distributed computation over fixed networks~\cite{uribe2018distributed}, accelerated primal--dual stochastic methods for regularized Wasserstein barycenters~\cite{dvurechensky2018decentralize}, quantized distributed optimization for semi-discrete entropic barycenters~\cite{krawtschenko2020distributed}, and time-varying-network formulations~\cite{yufereva2024decentralized}. These methods are important, but they operate through general optimization or dual-friendly formulations rather than by directly decentralizing the Sinkhorn geometric-mean step. Server-orchestrated optimal-transport/federated methods~\cite{rakotomamonjy2024federated,wei2025personalized} likewise retain a coordinator, whereas the present method implements the barycenter update with only neighbor-to-neighbor log-gossip and explicit trigger/quantization controls.


\section{Preliminaries}
\subsection{Problem data and notation}
We consider $N$ agents, each with a discrete probability $\mu_i\in\Delta_d$ on a common support $X=\{x_1,\dots,x_d\}$. Let $C\in\mathbb{R}^{d\times d}_{\ge 0}$ be the ground cost and, for $\varepsilon>0$, define the strictly positive Gibbs kernel
\begin{equation*}
K \;=\; \exp(-C/\varepsilon),
\end{equation*}
with element-wise $\exp(\cdot)$. Vectors are columns; $\odot$ and $\oslash$ denote element-wise product/division, $\mathbf 1$ is all-ones, and $\mathrm{softmax}(z)=\exp(z)/\langle \mathbf 1,\exp(z)\rangle$. We use $\|\cdot\|_1$ and $\|\cdot\|_\infty$. Table~\ref{tab:notation} summarizes all the notation and parameters used throughout this work.

\subsection{Centralized entropic barycenter (reference)}
Let $b^\star_\varepsilon$ denote the uniform-weight entropic Wasserstein barycenter of $\{\mu_i\}_{i=1}^N$ on $X$. The standard IBP/Sinkhorn barycenter keeps per-agent scalings $u_i\in\mathbb{R}^d_{>0}$ and a shared scaling $v\in\mathbb{R}^d_{>0}$. Given $v$, each agent updates
\begin{equation}
u_i \;\leftarrow\; \mu_i \,\oslash\, \big(K\,v + \eta\big), \qquad i=1,\dots,N ,
\label{eq:ui}
\end{equation}
with a small $\eta>0$ for numerical stability. The shared update is the geometric mean of $\{K^\top u_i\}$:
\begin{equation}
v \;\leftarrow\; \Big(\,\prod_{i=1}^N K^\top u_i \Big)^{\!1/N}.
\label{eq:v}
\end{equation}
A primal iterate is $b=v/\langle\mathbf 1,v\rangle=\mathrm{softmax}(\log v)$.

\subsection{Log-domain messages and averaging identity}
Define the per-agent log-message
\begin{equation}
s_i \;:=\; \log\!\big(K^\top u_i\big)\in\mathbb{R}^d. \label{eq:s}
\end{equation}
Then \eqref{eq:v} is equivalent to an arithmetic mean:
\[
\log v \;=\; \frac{1}{N}\sum_{i=1}^N s_i \quad\Longleftrightarrow\quad v=\exp\!\Big(\frac{1}{N}\sum_i s_i\Big).
\]
This identity is what we decentralize via local averaging.


\begin{table}[t]
\centering
\caption{Notation and parameters.}
\label{tab:notation}
\begin{tabular}{@{}ll@{}}
\toprule
\textbf{Symbol} & \textbf{Description} \\
\midrule
$N$ & Number of agents \\
$d$ & Support dimension \\
$C\in\mathbb{R}_{\geq 0}^{d\times d}$ & Ground cost matrix \\
$\varepsilon>0$ & Entropic regularization \\
$K=\exp(-C/\varepsilon)$ & Gibbs kernel \\
$\eta>0$ & Ridge parameter \\
$u_i\in\mathbb{R}_{>0}^d$ & Agent $i$ scaling vector \\
$v\in\mathbb{R}_{>0}^d$ & Shared scaling; $b=v/\langle \mathbf{1}, v\rangle$ \\
$s_i=\log(K^\top u_i)$ & Agent $i$ log-message \\
$z_i$ & Local estimate of $\frac{1}{N}\sum_j s_j$ \\
$\delta$ & Event-trigger threshold \\
$\tau_{\mathrm{in}}, \tau_{\mathrm{out}}$ & Inner/outer tolerances \\
$[s_{\min}, s_{\max}]$ & Message clipping range \\
$v_{\min}, v_{\max}$ & $e^{s_{\min}}, e^{s_{\max}}$ \\
$\Delta_q$ & Quantization error: $(s_{\max}-s_{\min})/2(2^b-1)$ \\
$L_{\exp}$ & Lipschitz const. of $\exp$: $e^{s_{\max}}$ \\
$L_{\mathrm{norm}}$ & Norm. map Lipschitz: $\leq 2/v_{\min}$ \\
$\rho\in(0,1)$ & Contraction factor \\
\bottomrule
\end{tabular}
\end{table}

\section{Methodology}
\label{sec:method}

We develop a fully decentralized variant of Sinkhorn's barycenter iteration by
(i) rewriting the shared geometric-mean update as an arithmetic average in the log-domain,
(ii) approximating that average via short rounds of neighborhood gossip, and
(iii) reducing bandwidth with event-triggered, quantized transmissions.
Throughout, vectors are columns; $\odot$ and $\oslash$ denote elementwise product/division; $\mathbf{1}$ is the all-ones vector; $\softmax(z) \coloneqq \exp(z)/\langle \mathbf{1},\exp(z)\rangle$.

\label{subsec:centralized}
Let $X=\{x_1,\dots,x_d\}$ be the common support and $C\in\mathbb{R}^{d\times d}_{\ge 0}$ the ground cost.
For $\varepsilon>0$ define the Gibbs kernel $K \coloneqq \exp(-C/\varepsilon)$ (element-wise exponential).
Each agent $i\in\{1,\dots,N\}$ holds a histogram $\mu_i\in\Delta^{d}$.
The entropic barycenter IBP/Sinkhorn iteration keeps per-agent scalings $u_i\in\mathbb{R}^d_{>0}$ and a shared scaling $v\in\mathbb{R}^d_{>0}$.
Given a small ridge $\eta>0$ for numerical stability, the updates are
\begin{align}
u_i &\leftarrow \mu_i \oslash \big(K v + \eta\mathbf{1}\big), \quad i=1,\dots,N, \label{eq:ibp-u} \\
v &\leftarrow \Big(\prod_{i=1}^N K^\top u_i\Big)^{\!1/N}. \label{eq:ibp-v}
\end{align}
Define the per-agent \emph{log-message}
\begin{equation}
s_i \;\coloneqq\; \log\!\big(K^\top u_i\big)\in\mathbb{R}^d. \label{eq:def-s}
\end{equation}
Then the shared update \eqref{eq:ibp-v} is equivalent to an arithmetic mean in the log-domain:
\begin{equation}
\log v \;=\; \frac{1}{N}\sum_{i=1}^N s_i \quad \Longleftrightarrow \quad v \;=\; \exp\!\Big(\tfrac{1}{N}\sum_{i=1}^N s_i\Big). \label{eq:log-id}
\end{equation}

\label{subsec:gossip}
Let $G=(V,E)$ be a connected, undirected communication graph with $|V|=N$ and neighbor set $\mathcal{N}_i$.
To avoid centralized aggregation in \eqref{eq:log-id}, each node $i$ maintains a local estimator $z_i \approx \tfrac{1}{N}\sum_j s_j$ and runs weighted gossip using the most recent neighbor packets $\tilde z_k$:
\begin{align}
z_i^{(0)} &\leftarrow s_i, \\
z_i^{(s+1)} &\leftarrow \sum_{k\in\mathcal{N}_i\cup\{i\}} w^{(s)}_{ik}\,\tilde z_k^{(s)}, \quad s=0,1,\dots \label{eq:gossip}
\end{align}
with (possibly time-varying) doubly-stochastic weights $w^{(s)}_{ik}$.
After a short inner consensus (see \S\ref{subsec:stopping}), each node forms a shared iterate
\begin{equation}
v \leftarrow \exp(z_i), \qquad b \leftarrow \softmax(\log v). \label{eq:shared-update}
\end{equation}
Interleaving \eqref{eq:ibp-u}, \eqref{eq:def-s}, \eqref{eq:gossip}, and \eqref{eq:shared-update} reproduces the centralized map without a coordinator.

\label{subsec:comms}
To minimize communication costs, node $i$ transmits only when its log-message changes appreciably.
Let $s_{i,\text{last}}$ be the last transmitted $s_i$.
On iteration $t$, transmit iff
\begin{equation}
\big\|s_i - s_{i,\text{last}}\big\|_\infty > \delta, \label{eq:trigger}
\end{equation}
Otherwise, neighbors reuse cached packets.
We communicate a $b$-bit \emph{quantized} and \emph{clipped} log-message.
Let $[s_{\min}, s_{\max}]$ bound the transmitted entries (locals remain full precision), and define a uniform $b$-bit quantizer over this interval.
The per-entry worst-case quantization error then satisfies
\begin{equation}
\Delta_q \;=\; \frac{s_{\max}-s_{\min}}{2\,(2^b-1)}. \label{eq:quant-error}
\end{equation}
Clipping applies \emph{only} to the packetized log-values; all local computations use the unclipped $s_i$.

\label{subsec:stopping}
We compute in the log-domain, keep $\eta>0$ in \eqref{eq:ibp-u}, and form the primal iterate via $b=\softmax(\log v)$.
For the inner consensus, node $i$ declares local agreement when
\begin{equation}
\max_{k\in\mathcal{N}_i}\big\|z_i-\tilde z_k\big\|_\infty < \tau,
\end{equation}
or after a fixed step cap.
For the outer loop, we stop when successive shared iterates stabilize, e.g.,
\begin{equation}
\big\|\log v^{(t+1)} - \log v^{(t)}\big\|_\infty < \tau.
\end{equation}

\subsection{Decentralized Sinkhorn: full procedure}
\label{subsec:algorithm}
\begin{algorithm}[t]
\caption{Decentralized log-gossip Sinkhorn with event triggering and quantization}
\label{alg:decent-sinkhorn}
\begin{algorithmic}[1]
\Require Graph $G$, kernel $K$, tolerances $\tau$, trigger $\delta$, quantizer $\mathcal{Q}_b(\cdot)$, clip range $[s_{\min},s_{\max}]$, ridge $\eta>0$.
\For{each node $i\in V$ \textbf{in parallel}}
  \State $u_i \gets \mathbf{1}$;\quad $s_i \gets \log(K^\top u_i)$;\quad $z_i \gets s_i$;\quad $s_{i,\text{last}} \gets s_i$
\EndFor
\While{not converged}
  \For{each node $i$ \textbf{in parallel}} \Comment{Local scaling}
    \State $v \gets \exp(z_i)$;\quad $u_i \gets \mu_i \oslash (K v + \eta\mathbf{1})$ \label{line:local}
    \State $s_i \gets \log(K^\top u_i)$ \label{line:recompute-s}
    \If{$\|s_i - s_{i,\text{last}}\|_\infty > \delta$} \Comment{Event trigger}
      \State send $\mathcal{Q}_b\!\big(\mathrm{clip}(s_i;[s_{\min},s_{\max}])\big)$ to $\mathcal{N}_i$;\; $s_{i,\text{last}} \gets s_i$
    \EndIf
  \EndFor
  \State \textbf{(Inner gossip)} Each node $i$ updates $z_i$ using \eqref{eq:gossip} and most recent neighbor packets until $\max_{k\in\mathcal{N}_i}\|z_i-\tilde z_k\|_\infty<\tau$ or a step cap is reached.
  \For{each node $i$ \textbf{in parallel}} \Comment{Shared projection}
    \State $v \gets \exp(z_i)$;\quad $b \gets \softmax(\log v)$
  \EndFor
\EndWhile
\State \Return $b$
\end{algorithmic}
\end{algorithm}

\section{Theoretical Results}
We analyze the decentralized, event–triggered, quantized log–gossip Sinkhorn scheme. Our goals are to: (i) establish a single, consistent contraction for the centralized IBP/Sinkhorn \emph{barycenter} map in the Hilbert metric; (ii) quantify inner-loop consensus time via the spectral gap; (iii) propagate consensus/trigger/quantization perturbations through the $\exp$+normalize pipeline in $\ell_1$ with explicit, range-aware constants; and (iv) state communication/compute complexity with consistent $d$-dependence and robust handling of asynchrony.

\subsection{Assumptions, stabilization, and range management}
\textbf{Assumption 1 (Network \& weights).}
The communication graph $G=(V,E)$ is connected. Averaging weights $W=[w_{ik}]$ are doubly stochastic with $w_{ik}\ge \beta>0$ for $(i,k)\in E\cup\{(i,i)\}$. For time-varying/asynchronous updates we specify assumptions in \S\ref{sec:async}.

\smallskip
\textbf{Assumption 2 (Problem regularity \& stabilization).}
The Gibbs kernel $K=\exp(-C/\varepsilon)$ has strictly positive entries for $\varepsilon>0$. Each agent $i$ holds a histogram $\mu_i\in\Delta^d$. We run stabilized IBP updates with \emph{positive} ridge $\eta>0$ in the $u$-update (cf.\ (1)) so that $K^\top u_i+\eta\mathbf 1>0$ always. This implies that the per-agent log-message
\[
s_i := \log\!\big(K^\top u_i\big)
\]
is well-defined at every iteration even if $\mu_i$ is sparse (contains zeros).

\smallskip
\textbf{Assumption 3 (Finite log-range via clipping).}
There exist finite bounds $s_{\min}\le s_{\max}$ such that \emph{transmitted} log-messages satisfy $s_i^{\text{tx}}=\mathrm{clip}(s_i;[s_{\min},s_{\max}])\in[s_{\min},s_{\max}]^d$ before quantization. Define $v_{\min}:=e^{s_{\min}},\;v_{\max}:=e^{s_{\max}}$. We quantize $s_i^{\text{tx}}$ componentwise with a $b$-bit uniform quantizer over $[s_{\min},s_{\max}]$, yielding per-entry error $\Delta_q=(s_{\max}-s_{\min})/2(2^b-1)$. Receivers cache the latest packet and reuse it until a new one arrives.

\begin{remark}[Why clipping is part of the algorithm]
Clipping makes the analysis self-consistent and removes pathological range blow-ups at small $\varepsilon$ or for peaky histograms. It is applied \emph{only to the communicated log-message}; local computations retain full precision. If adaptive ranges are used, nodes can piggyback current $(s_{\min},s_{\max})$ with occasional offset/scale updates.
\end{remark}

\renewcommand{\oscK}{\operatorname{osc}(\log K)}
\renewcommand{\dH}{d_H}

\subsection{Barycenter-map contraction in the Hilbert metric}
Let $F:\Delta^{d-1}\!\to\!\Delta^{d-1}$ denote one \emph{full} barycenter cycle:
all $u_i$-updates with a common $v$, followed by the shared $v$-update and
normalization $b=v/\langle\mathbf{1},v\rangle$. We use the Hilbert projective
metric $\dH(x,y)=\log\!\big(\max_j x_j/y_j\big)-\log\!\big(\min_j x_j/y_j\big)$
on $\mathbb{R}^d_{>0}$.

\begin{lemma}[Hilbert contraction (centralized)]\label{lem:hilbert}
Let $\mathrm{osc}(\log K)=\max_{j,j'}\max_\ell(\log K_{\ell j}-\log K_{\ell j'})$.
Define
\begin{equation}\label{eq:theta-rho-def}
\theta \coloneqq \tanh\!\Big(\tfrac{1}{4}\,\oscK\Big), \qquad
\rho \coloneqq \theta^{2}.
\end{equation}
Then
\begin{equation}\label{eq:hilbert-contraction}
\dH\!\big(F(b),F(b')\big) \le \rho\, \dH(b,b').
\end{equation}
Moreover,
\begin{equation}\label{eq:rho-bound}
\rho \le \tanh^{2}\!\Big(\tfrac{\|C\|_\infty}{2\varepsilon}\Big) < 1,
\end{equation}
so $\rho \uparrow 1$ as $\varepsilon \downarrow 0$.
\end{lemma}

\noindent\emph{Sketch.}
Each $u_i$-update and the shared $v$-update are induced by positive linear maps followed by projective normalization. The classical Birkhoff–Bushell theory gives a per-map coefficient $\le\theta$ \cite{lemmens2012nonlinear}. One full IBP barycenter cycle composes two such positive normalized maps, thus contracting with factor at most $\theta^2$, independent of $N$. \hfill\qed

\subsection{Bridging $\dH$ and $\ell_1$ under range bounds}
\begin{lemma}[Hilbert$\to\ell_1$ bridge]\label{lem:bridge}
If $x,y\in\mathbb{R}^d_{>0}$ satisfy $v_{\min}\le x_j,y_j\le v_{\max}$ and
$p=\proj{x}$, $q=\proj{y}$, then
\begin{equation}\label{eq:bridge-l1}
\|p-q\|_1 \le \frac{2}{v_{\min}}\,\|x-y\|_1 .
\end{equation}
Moreover, if $\dH(x,y)\le \gamma$, then
\begin{equation}\label{eq:bridge-hilbert}
\|p-q\|_1 \le \frac{2}{v_{\min}}\,(e^{\gamma}-1).
\end{equation}
\end{lemma}

\subsection{One-step perturbation through \texorpdfstring{$\exp$}{exp} and normalization}
Let $\widetilde F$ be the decentralized counterpart of $F$ that uses inner-loop consensus with tolerance $\tau$, event-trigger threshold $\delta$, and $b$-bit quantization with step $\Delta_q$ on the \emph{transmitted} log-messages.

\begin{lemma}[One-step perturbation]\label{lem:onestep}
Let $L_{\exp}:=e^{s_{\max}}$ and
\[
L_{\mathrm{norm}}
:=\sup_{x>0}\Big\|\frac{x}{\langle\mathbf 1,x\rangle}-\frac{\cdot}{\langle\mathbf 1,\cdot\rangle}\Big\|_{1\leftarrow 1}
\;\;\text{(operator $\ell_1\!\to\!\ell_1$ norm)}.
\]
Then for any $b\in\Delta^{d-1}$,
\[
\|\widetilde F(b)-F(b)\|_{1}\;\le\;L_{\exp}\,L_{\mathrm{norm}}\;(\tau+\delta+\Delta_q).
\]
Under Assumption~3 (finite log-range), $L_{\mathrm{norm}}\le 2/v_{\min}$, hence
\[
\|\widetilde F(b)-F(b)\|_{1}\;\le\;\frac{2\,e^{s_{\max}}}{v_{\min}}\;(\tau+\delta+\Delta_q).
\]
\end{lemma}

\noindent\emph{Discussion.}
The $(\tau+\delta+\Delta_q)$ term is a \emph{worst-case} $\ell_\infty$ bound per coordinate on the pre-$\exp$ log-vector used by each node. No independence between error sources is assumed; we simply add magnitudes to cover correlation/staleness in caches.

\subsection{Tracking the centralized barycenter and $\varepsilon$-sensitivity}
Let $b^\star_\varepsilon$ be the centralized entropic barycenter fixed point.

\begin{theorem}[Linear convergence to a neighborhood of $b^\star_\varepsilon$]\label{thm:track}
Under Assumptions~1–3 and Lemmas~\ref{lem:hilbert}–\ref{lem:onestep}, the decentralized iterates obey
\begin{equation}
\big\|\widetilde b^{(t)}-b^\star_\varepsilon\big\|_1
\;\le\;
\rho^{\,t}\,\big\|\widetilde b^{(0)}-b^\star_\varepsilon\big\|_1
\;+\;
\frac{L_{\exp}L_{\mathrm{norm}}}{1-\rho}\,(\tau+\delta+\Delta_q),
\end{equation}
with $\rho$ from Lemma~\ref{lem:hilbert}.
\end{theorem}

\begin{remark}
Because $\rho\le \tanh^2(\|C\|_\infty/(2\varepsilon))$, the rate degrades as $\varepsilon\downarrow 0$. Moreover $L_{\exp}/v_{\min}=e^{s_{\max}-s_{\min}}$ can grow when ranges widen at small $\varepsilon$. Our analysis therefore recommends either (i) continuation in $\varepsilon$ or (ii) tighter clipping ranges for small $\varepsilon$.
\end{remark}

\subsection{Inner-loop consensus time}
Let $z_i^{(s)}$ denote node $i$'s inner gossip state after $s$ steps, initialized at $z_i^{(0)}=s_i^{\text{tx}}$. For the synchronous case $W^{(s)}\equiv W$:

\begin{lemma}[Consensus via spectral gap]\label{lem:consensus}
If $\sigma_2(W)$ is the second largest singular value, then
\begin{equation}\label{eq:consensus-decay}
\norm{z^{(s)}-\bar z\,\mathbf 1}_2
\;\le\;
\sigma_2(W)^{\,s}\,\norm{z^{(0)}-\bar z\,\mathbf 1}_2.
\end{equation}
Consequently, the inner steps needed for tolerance $\tau$ satisfy
\begin{equation}\label{eq:Tinner}
T_{\mathrm{inner}}(\tau)
=\mathcal{O}\!\left(
\frac{\log\!\big(\sqrt d\,\norm{z^{(0)}-\bar z\,\mathbf 1}_2/\tau\big)}
{1-\sigma_2(W)}
\right).
\end{equation}
If $W$ are Metropolis (or similar) weights on an undirected graph,
then $1-\sigma_2(W)\gtrsim \lambda_2(G)/d_{\max}$.
\end{lemma}

\begin{proposition}[Trigger rate from a variation budget]\label{prop:trigger}
Under Theorem~\ref{thm:track}, there exist constants $c_1,c_2$ depending only on $(s_{\min},s_{\max})$ and problem data such that
\begin{equation}
V_i(T)\;\le\;c_1\,\rho^{\,T}\;+\;c_2\,\frac{1-\rho^{\,T}}{1-\rho}\,(\tau+\delta+\Delta_q).
\end{equation}
Consequently, the number of broadcasts by node $i$ over $T$ outer iterations satisfies
\begin{equation}
M_i(T)\;\le\;1+\Big\lceil V_i(T)/\delta\Big\rceil.
\end{equation}
\end{proposition}

\noindent\emph{Sketch.}
One-step contraction contributes the $\rho^t$ decay; the steady-state bias from Lemma~\ref{lem:onestep} yields the geometric series; the trigger converts cumulative variation into event counts.

\subsection{Communication and computation complexity}
Let $\Touter$ be the number of outer IBP cycles and $\Tinner$ the number of
inner gossip steps per outer iteration. Choose $(\tau,\delta,\Delta_q)$ so that
\begin{equation}\label{eq:neighborhood}
\frac{L_{\exp}L_{\mathrm{norm}}}{1-\rho}\,(\tau+\delta+\Delta_q)\ \le\ \tfrac{1}{2}\,\varepsilon_{\mathrm{tar}}.
\end{equation}
To reach $\|\widetilde b^{(T)}-b^\star_\varepsilon\|_1\le \varepsilon_{\mathrm{tar}}$, it suffices to take
\begin{equation}\label{eq:Touter}
\Touter=\mathcal{O}\!\left(
\frac{\log(1/\varepsilon_{\mathrm{tar}})}{\log(1/\rho)}
\right),
\end{equation}
and
\begin{equation}\label{eq:Tinner-comm}
\Tinner=\mathcal{O}\!\left(
\frac{\log(d/\tau)}{1-\sigma_2(W)}
\right).
\end{equation}

\noindent The message counts satisfy
\begin{equation}\label{eq:Mcounts}
\begin{aligned}
M_{\mathrm{total}} &= \mathcal{O}(|E|\,T), \quad
M_{\mathrm{agent}} = \mathcal{O}(\deg(G)\,T), \\
T &= T_{\mathrm{outer}} T_{\mathrm{inner}}.
\end{aligned}
\end{equation}
Per outer iteration and per agent, computing $K^\top u_i$ costs
$\mathcal{O}(d^2)$ (or $\mathcal{O}(d\log d)$ with fast transforms), while
serialization per inner step costs $\mathcal{O}(\deg(G)\cdot d)$. Writing
$\kappa(d)\in\{d^2,\,d\log d\}$, the total runtime across all agents is
\begin{equation}\label{eq:runtime}
\mathcal{O}\!\Big(\Touter\big(N\,\kappa(d)+\Tinner\,|E|\,d\big)\Big).
\end{equation}
Convergence is linear per iteration (exponential in the count) with factor $\rho$.

\subsection{Asynchrony, delays, and stale packets}\label{sec:async}
We now replace the synchronous inner loop with a standard randomized/asynchronous model.
Let $\{W^{(s)}\}$ be i.i.d.\ doubly stochastic averaging matrices with $\mathbb{E}[W^{(s)}]=\overline W$ and spectral gap $1-\sigma_2(\overline W)>0$. Assume bounded communication delays with staleness of at most $\Delta$ inner steps and cached packets, and that the \emph{activation process} is independent of the quantization noise. Then:

\begin{theorem}[Asynchronous consensus in expectation]\label{thm:async}
Under the model above,
\begin{equation}
\mathbb{E}\|z^{(s)}-\bar z\,\mathbf 1\|_2 \;\le\; \sigma_2(\overline W)^{\,s}\,\mathbb{E}\|z^{(0)}-\bar z\,\mathbf 1\|_2,
\end{equation}
so the bound in Lemma~\ref{lem:consensus} holds \emph{in expectation} with $1-\sigma_2(W)$ replaced by $1-\sigma_2(\overline W)$. Bounded staleness simply enlarges the effective trigger/cache term, which is already accounted for by $(\delta,\tau)$ in Lemma~\ref{lem:onestep} and Proposition~\ref{prop:trigger}.
\end{theorem}

\noindent\emph{Remark (Notation and symmetry).}
When $W$ is symmetric, $\sigma_2(W)=\lambda_2(W)$ is the second largest eigenvalue in magnitude. We consistently use $\|\cdot\|_{1\leftarrow 1}$ for operator $\ell_1$-norms.

\medskip
\noindent\emph{Summary.}
A uniform Hilbert contraction for the centralized (Lemma~\ref{lem:hilbert}), spectral-gap inner consensus (Lemmas~\ref{lem:consensus}, \ref{thm:async}), and range-aware perturbation propagation (Lemma~\ref{lem:onestep}) together yield linear tracking with an explicit steady-state neighborhood scaling linearly in $(\tau,\delta,\Delta_q)$ (Theorem~\ref{thm:track}), and message/runtime complexity bounds.


\section{Results}\label{sec:results}

We evaluated the proposed decentralized, event-triggered log-gossip algorithm. Experiments use a \(4\times 4\) grid (\(N{=}16\)), support size \(d{=}64\), entropic parameter \(\varepsilon\), and inner consensus tolerance \(\tau\).
For each condition, we average over multiple seeds and report 95\% confidence intervals.
Accuracy is measured against the centralized entropic barycenter.

\begin{figure}[t]
   \centering
  \includegraphics[width=\linewidth]{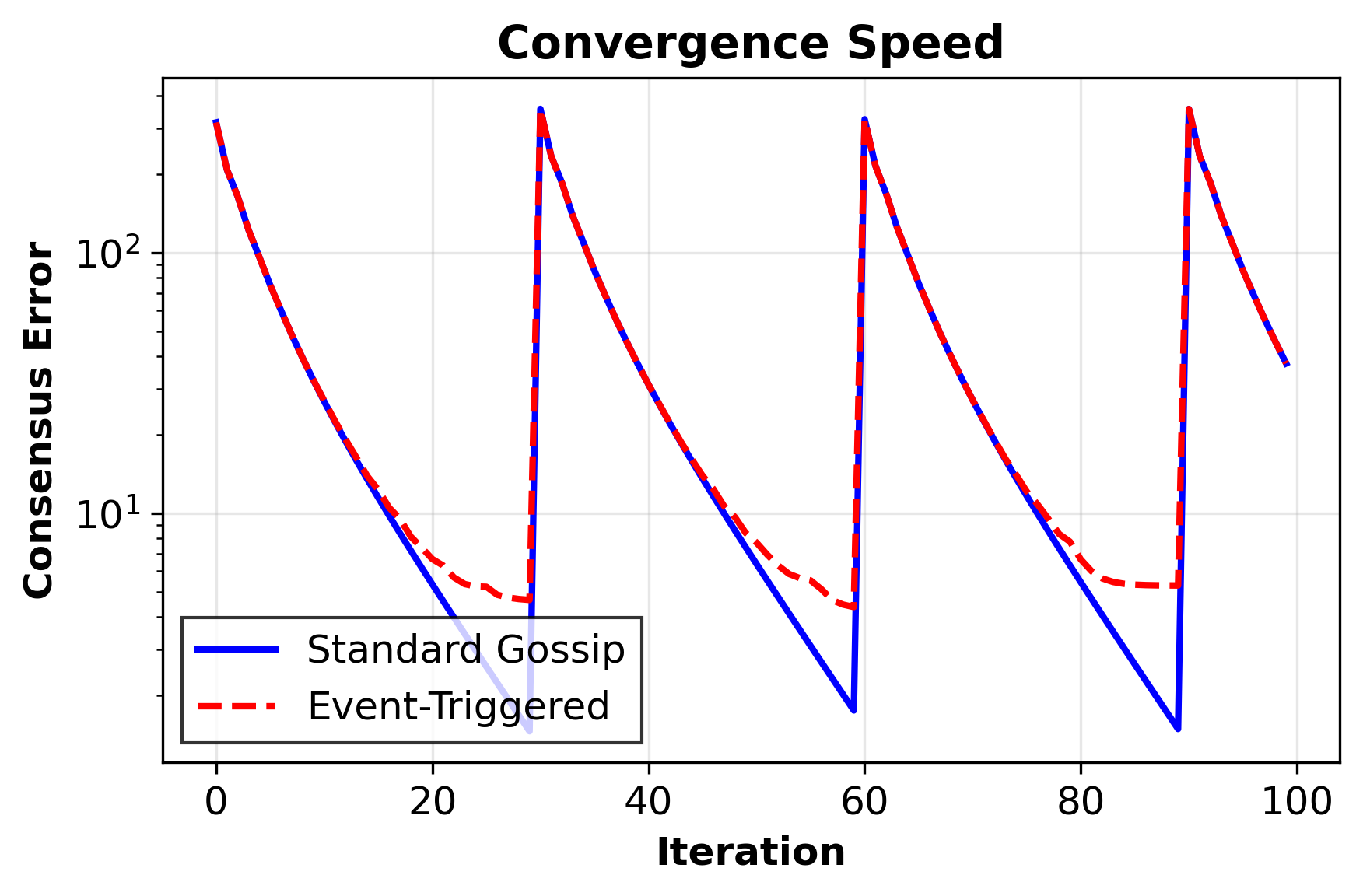}
  \caption{Convergence traces (log residual) for always-gossip vs event-triggered.}
  \label{fig:conv}
\end{figure}

\begin{figure}[t]
  \centering
  \includegraphics[width=\linewidth]{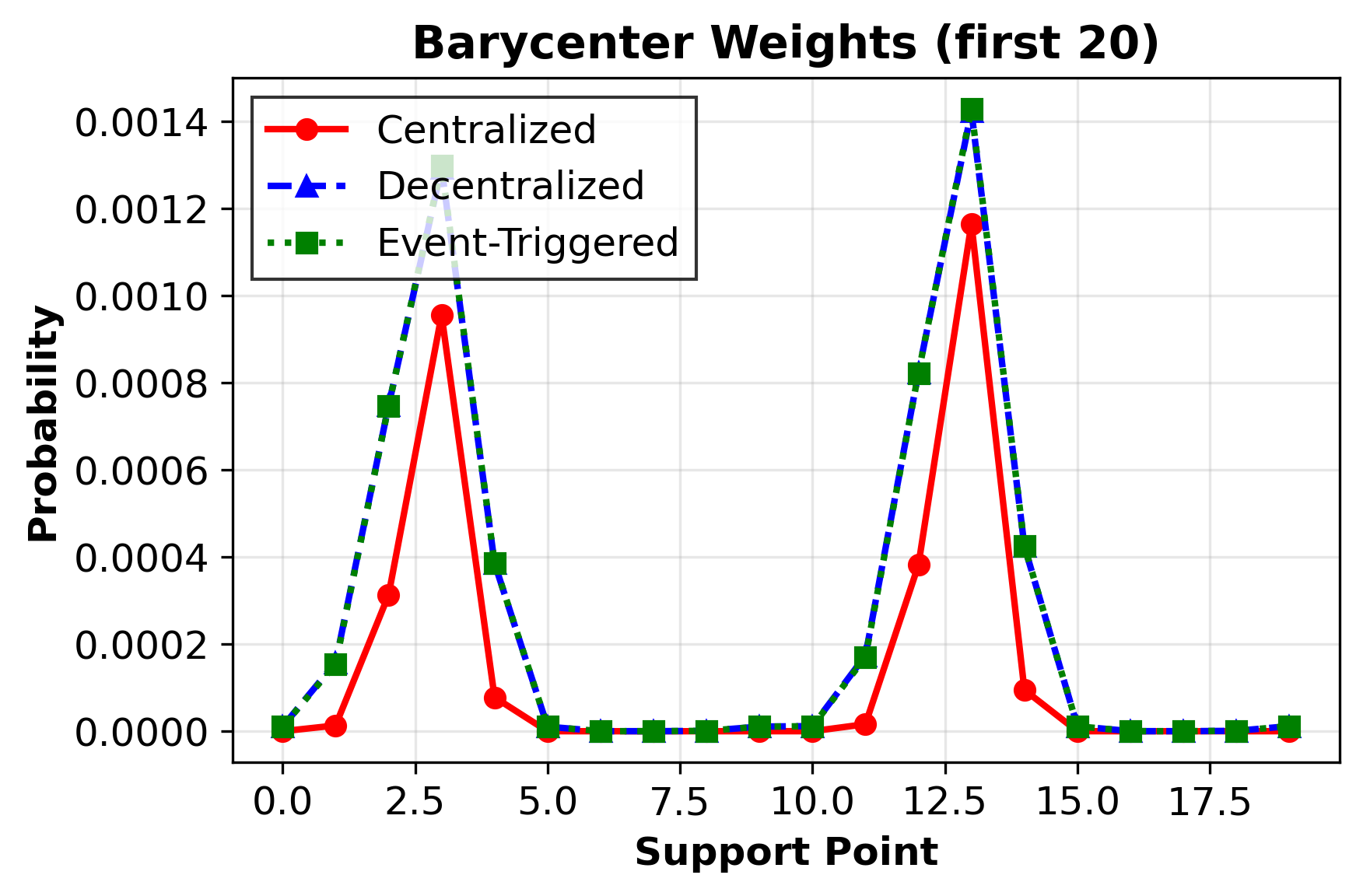}
 \caption{Centralized vs decentralized barycenter overlap on the common support.}
  \label{fig:overlap}
\end{figure}

Figure~\ref{fig:conv} plots the inner loop consensus residual on a logarithmic scale for ``always gossip'' and the event-triggered variant.
Both traces exhibit the expected \emph{sawtooth} profile: rapid inner contraction inside each outer iteration, followed by a small jump at the next outer update, which is consistent with the spectral-gap–controlled consensus rate (cf. Lemmas on gossip contraction).
Event triggering slows the inner convergence modestly but reaches the same residual floor, which matches the theory that triggering introduces bounded perturbations without changing the contraction regime. Figure~\ref{fig:overlap} overlays the centralized and decentralized barycenters on the common support.
The curves are visually indistinguishable over the dominant modes, indicating that the decentralized iterate remains in a small \(\ell_1\)-neighborhood of the centralized solution.
This observation is consistent with the tracking bound of Theorem~\ref{thm:track}

\begin{figure}[t]
  \centering
  \includegraphics[width=\linewidth]{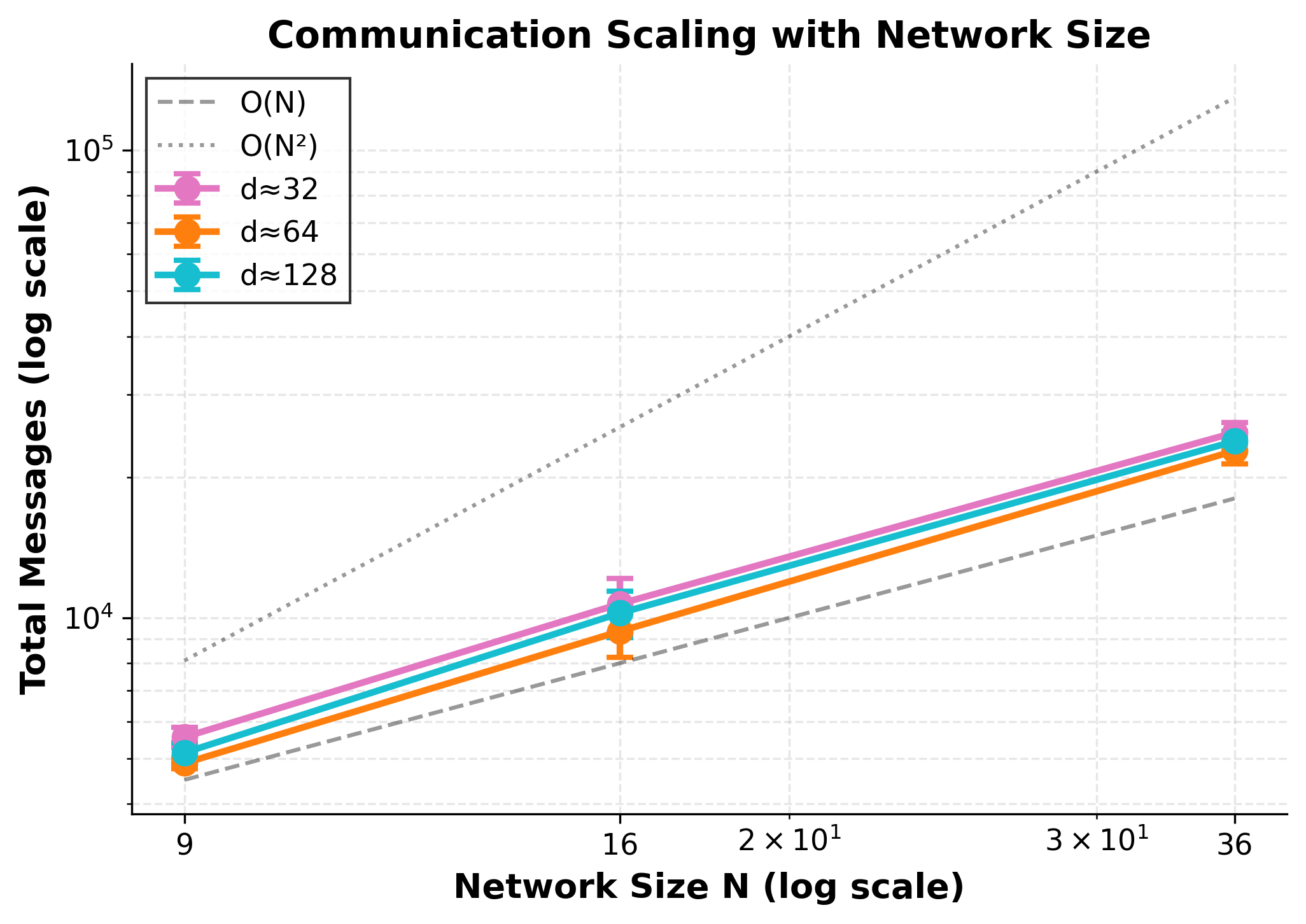}
  \caption{Total messages vs network size (log--log) with $O(N)$ and $O(N^2)$ guides.}
  \label{fig:scaling-comm}
\end{figure}

\begin{figure}[t]
  \centering
  \includegraphics[width=\linewidth]{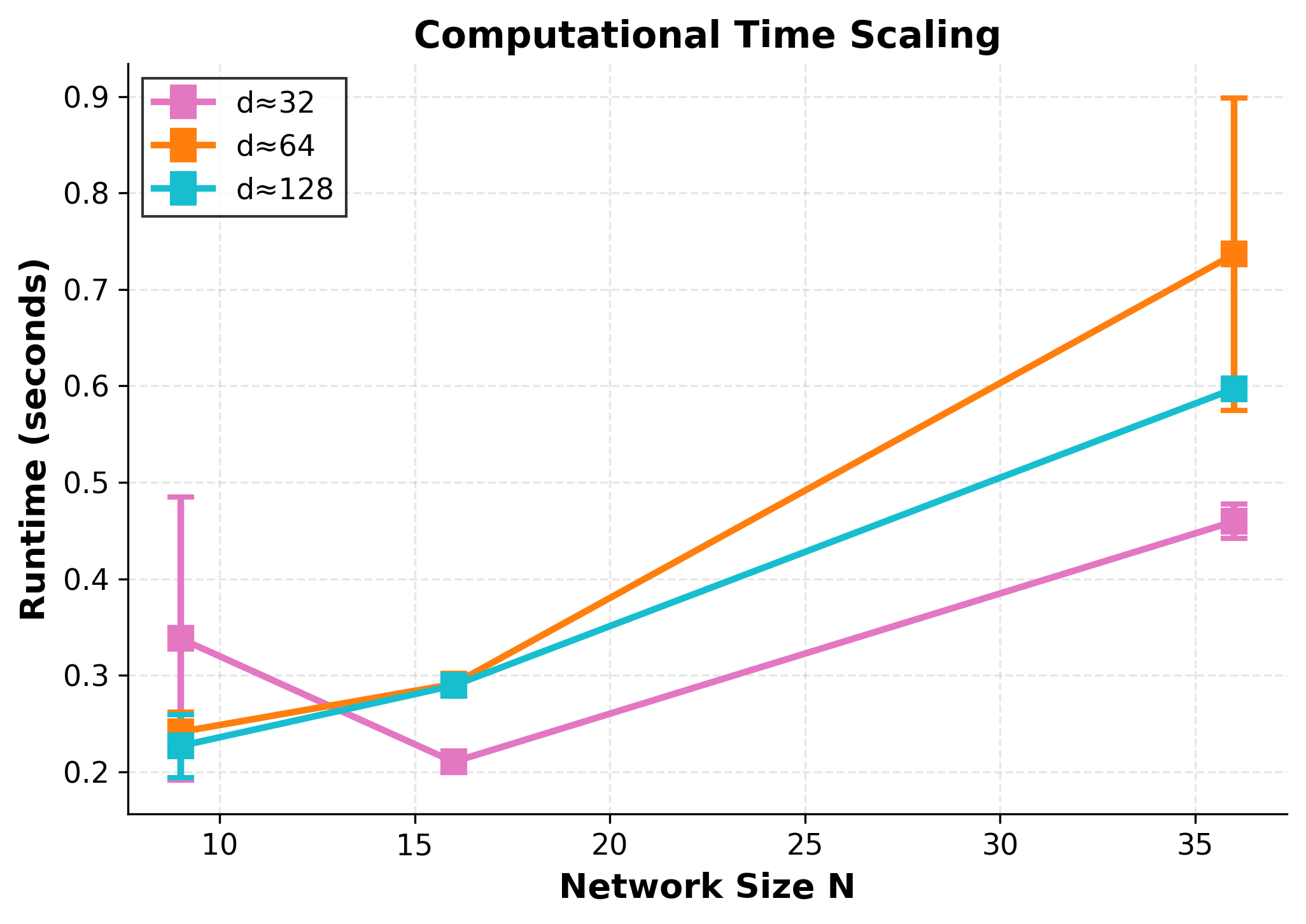}
  \caption{Runtime vs network size.}
  \label{fig:runtime}
\end{figure}

\begin{figure}[t]
 \centering
  \includegraphics[width=\linewidth]{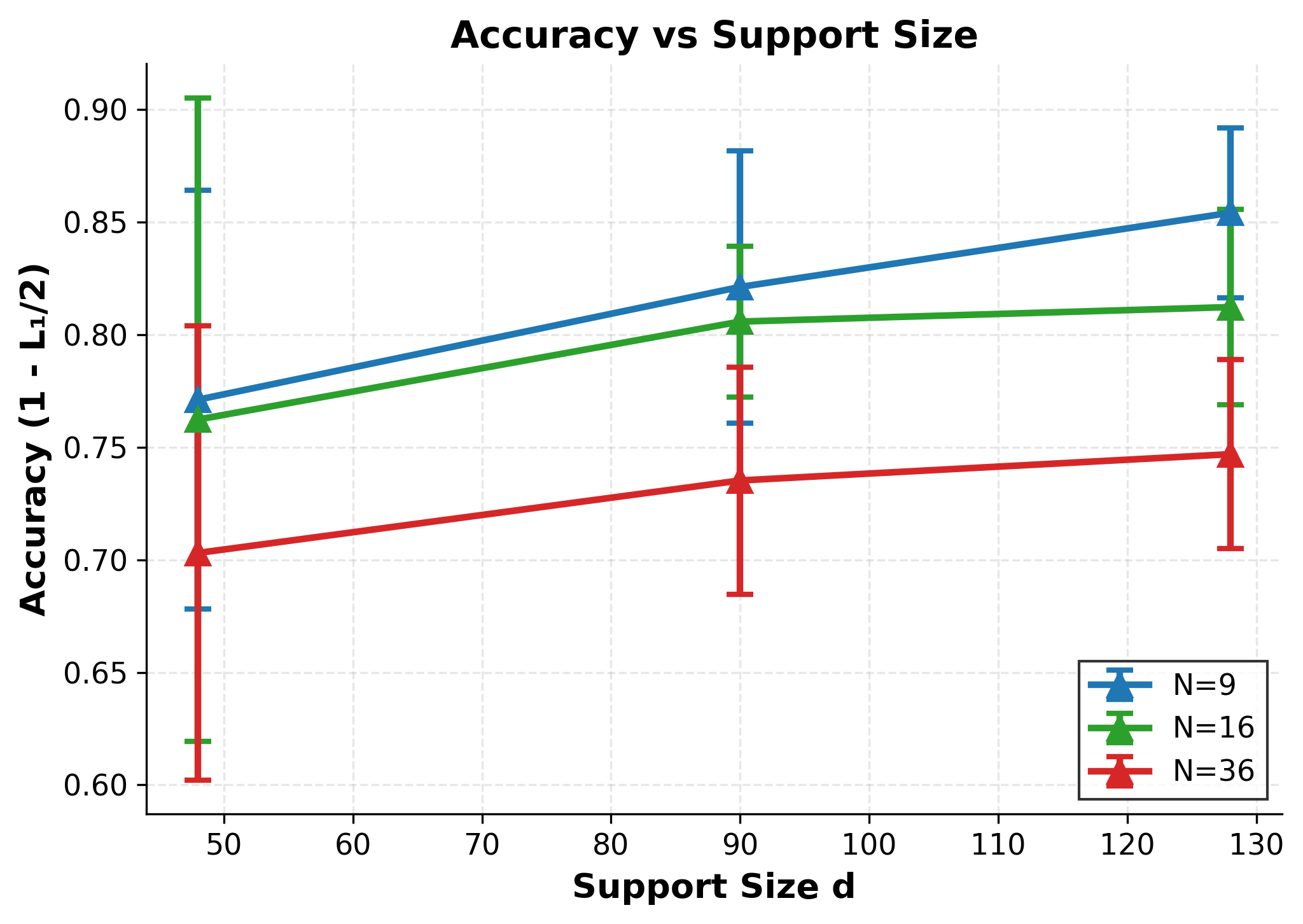}
  \caption{Accuracy vs support size $d$.}
  \label{fig:support}
\end{figure}

Figure~\ref{fig:scaling-comm} shows the total message count versus network size on a log–log scale with \(O(N)\) and \(O(N^2)\) reference lines.
In grid graphs, the empirical slope lies close to \(O(N)\) when \(\tau\) and the inner step cap are fixed, in line with the observation that (i) the number of inner gossip steps per outer iteration remains roughly constant and (ii) the messages scale with the number of edges \(|E|\!\sim\!O(N)\).
Figure~\ref{fig:runtime} reports wall-clock runtime, which increases monotonically with \(N\) as expected from the per-iteration cost analysis. Figure~\ref{fig:support} plots the accuracy versus the support size \(d\).
Accuracy improves with \(d\) and then saturates, matching the expected trade-off between the discretization error and the entropic bias at fixed \(\varepsilon\).

\vspace{-3 mm}
\section{CONCLUSIONS}

We proposed a decentralized Sinkhorn algorithm that recasts the centralized geometric mean as a log-domain arithmetic average approximated via local gossip, eliminating central coordination while preserving the Bregman projection structure. Event-triggered transmissions and quantization provide tunable accuracy-bandwidth trade-offs that tolerate asynchrony and packet loss. Theoretical analysis established linear convergence to a neighborhood of the centralized barycenter, with bias scaling linearly with consensus tolerance, trigger threshold, and quantization step. Experiments confirmed near-centralized accuracy with substantially fewer messages. Future directions include heterogeneous supports, adaptive quantization, and real-world deployment in distributed learning or sensor fusion.

\vspace{-4 mm}
\bibliographystyle{unsrt}
\bibliography{ACC2026}

\end{document}